\documentclass[aps,pra,twocolumn,showpacs,superscriptaddress,a4paper,groupedaddress]{revtex4-1}
\usepackage{graphicx}    
\usepackage{dcolumn}    
\usepackage{bm}             
\usepackage{amssymb}   
\usepackage{amsmath}    
\usepackage{subfigure}    
\usepackage{braket}
\usepackage{mathrsfs}
\usepackage[export]{adjustbox}
\usepackage[toc,page]{appendix}
\begin{document}

\title{Coherent Forward Broadening in Cold Atom Clouds}
\author{R.T. Sutherland}
\email{rsutherl@purdue.edu}

\author{F. Robicheaux}
\email{robichf@purdue.edu}
\affiliation{Department of Physics and Astronomy, Purdue University, West Lafayette IN, 47907 USA}
\affiliation{Purdue Quantum Center, Purdue University, West Lafayette,
Indiana 47907, USA}

\date{\today}

\begin{abstract}

It is shown that homogeneous line-broadening in a diffuse cold atom cloud is proportional to the resonant optical depth of the cloud. Further, it is demonstrated how the strong directionality of the coherent interactions causes the cloud's spectra to depend strongly on its shape, even when the cloud is held at constant densities. These two numerical observations can be predicted analytically by extending the single photon wavefunction model. Lastly, elongating a cloud along the line of laser propagation causes the excitation probability distribution to deviate from the exponential decay predicted by the Beer-Lambert law to the extent where the atoms in the back of the cloud are more excited than the atoms in the front. These calculations are conducted at low densities relevant to recent experiments.

\end{abstract}
\pacs{ }
\maketitle

\section{Introduction}\label{sec:intro}
Since the seminal work of Dicke \cite{dicke1954}, the effects of collective emission in an ensemble of radiators has been studied extensively. Collective long-range interactions have shown their import in many phenomena such as coherent forward scattering \cite{rouabah2014,svidzinsky2008}, the collective Lamb shift \cite{meir2014}, fault-tolerant quantum computation \cite{yavuz2015}, laser cooling \cite{burnett1991}, and homogeneous line-broadening \cite{pellegrino2014}. The study of collective effects and their role in transition lines is of particular importance for the implementation of highly accurate atomic clocks \cite{ido2005, chang2004, katori2003, ludlow2015, shiga2009}, where achieving narrow resonance lines is essential. As the quest for extremely accurate atomic clocks progresses, an understanding of the plethora of physical processes in cold atomic gasses, such as density dependent line-broadening \cite{ido2005}, will need to be understood using models that extend the classical theories of line-broadening, since these models mainly rely on local interactions \cite{friedberg1973,baranger1958,breene1970}. 

Although the interaction between an individual pair of atoms or molecules in a cold, diffuse gas can be tiny, the long-range nature of the dipole-dipole couplings can lead to the substantial constructive buildup of small interactions over an entire ensemble \cite{svidzinsky2008,svidzinsky_chang2008,rouabah2014,bienaime2012,courteille2010,scully2007,javanainen2014}. This understanding has been improved using single photon wavefunction theories that provide analytic predictions about the line-broadening seen in an atomic cloud when driven by an extremely weak laser \cite{scully2007,svidzinsky2008, svidzinsky_chang2008}. Here we explore this concept for very low atomic density, showing that the model remains valid whenever inhomogeneous broadening is negligible. For example, the line-width ($\Gamma^{\prime}$) for an N atom gas with a Gaussian density distribution increases according to $\Gamma^{\prime} = (1 + \frac{\xi b_{0}}{8})\Gamma$, where $b_{0}$ is the cooperativity parameter, $b_{0}\equiv \frac{3(N-1)}{k^{2}\sigma^{2}}$, $\xi$ is a number parameterizing the cloud-shape (see Eq. \ref{eq:dens}), N is the number of atoms in the cloud, and $\Gamma$ is the single atom line-width. Unlike collisional broadening where the extra line-width is only proportional to the average density, the extra line-width from the dipole-dipole interaction is proportional to the average density \textit{and} the linear size of the gas. This scaling should affect the spectroscopy of cold atoms since the extra line width can be substantial, even at low densities. This also has implications for fault tolerant quantum computation, specifically the threshold theorem which assumes spatially and temporally local interactions \cite{terhal2005,hui2009}. Here large separations ensure the absence of collisional broadening but the dipole-dipole interaction will still lead to an extra line-width, implying a faster decoherence \cite{yavuz2015}.

\begin{figure}[!h]
	\includegraphics[width=0.50\textwidth,scale=0.1]{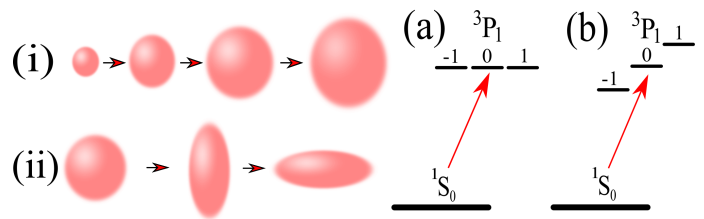}
	\label{fig:pretty}
    \caption{The scattered radiation is studied for (i) clouds increasing in number of atoms while being held at constant average density and shape, and (ii) clouds with varying shapes with respect to $\boldsymbol{\hat{k}}$ held at constant average density and atom number. The figure also shows the level diagram of the transitions focused on in this letter: the $^{88}$Sr J=0 to J=1 intercombination line where the $^{3}P_{1}$ level can be either degenerate in $M_{J}$ (a) or Zeeman split (b).}
\end{figure}

By studying the photon scattering rate versus detuning, we illustrate how cold atom clouds with average densities such that $\bar{\rho}/k^{3}\ll 1$, exhibit collective effects similar to those predicted using single photon wavefunction theories \cite{scully2007, svidzinsky2008, svidzinsky_chang2008}. Even at densities where individual interactions are tiny, a cloud can still show signs of coherence due to the buildup of the $\frac{e^{ikr}}{r}$ term present in dipole-dipole couplings. This leads the line-broadening and the excitation distribution of a cloud to depend strongly on its overall shape, since interactions add constructively between atoms separated by a position vector parallel to the direction of the driving laser ($\boldsymbol{\hat{k}}\equiv\frac{\boldsymbol{k}}{|\boldsymbol{k}|}$). For constant $\bar{\rho}$ and N, clouds that are elongated parallel to $\boldsymbol{\hat{k}}$ are more broadened than those elongated perpendicular to $\boldsymbol{\hat{k}}$. 

This work is organized as follows: In Sec. II, we describe our numerical approach, as well as a variant of the Gauss-Seidel iteration routine that allows for the simulation of larger numbers of atoms. In Sec. III, we provide an analytic derivation of the dependence of a cloud's line-shape on its density, aspect ratio, and number of atoms. This is done using an improvement of the single photon wavefunction model so that it includes a vectorized electromagnetic field, as well as a spherically non-symmetric Gaussian cloud. In Sec. IV, we discuss the results of our numerical model. Here we show that until a cloud becomes too elongated along the line of laser propagation, the results of our analytic model are numerically accurate. Lastly, we show that for clouds highly elongated along $\boldsymbol{\hat{k}}$ and driven by a red detuned laser, a counter-intuitive excitation distribution develops: the atoms in the back of the cloud (farther along $\boldsymbol{\hat{k}}$) are more likely to be excited than the atoms in the front of the cloud. This is the reverse of the typical exponential decay predicted by the Beer-Lambert law. We discuss experimental possibilities and provide concluding remarks in Sec. V.

\section{Numerical Method}\label{sec:method}

For a weak laser, a collection of atoms can be treated as classical radiating dipoles or equivalently as coupled damped harmonic oscillators \cite{javanainen2014, jenkins2012, svidzinsky2010, ruostekoski1997}, 
\begin{eqnarray}
\label{eq:a}
\dot{a}_{\alpha}^{\mu}(t)&=\nonumber&(i\Delta - \Gamma /2)a_{\alpha}^{\mu}(t) -  i(d /\hbar)E^{\mu}(\boldsymbol{r}_{\alpha}) \\ &-& (\Gamma /2)\sum_{\beta\neq\alpha,\nu}G_{\mu\nu}({\boldsymbol{r}}_{\beta} - {\boldsymbol{r}}_{\alpha})a_{\beta}^{\nu}(t),
\end{eqnarray}
where $a_{\alpha}^{\mu}$ represents the $\mu^{th}$ polarization amplitude of the $\alpha^{th}$ atom, d is the electric dipole matrix element, $E^{\mu}(\boldsymbol{r}_{\alpha}$) is the $\mu^{th}$ component of the laser field at atom $\alpha$, $\Delta$ is the detuning, and $G_{\mu\nu}(\boldsymbol{r})$ is the usual dipole field propagator \cite{jackson1999},
\begin{equation}
G_{\mu\nu}(\boldsymbol{r}) = \frac{3e^{ikr}}{2ikr} \{[\delta_{\mu\nu} - \hat{r}_{\mu}\hat{r}_{\nu}] + [\delta_{\mu\nu} - 3\hat{r}_{\mu}\hat{r}_{\nu}][\frac{i}{kr} - \frac{1}{(kr)^{2}}]\},
\label{eq:g}
\end{equation}
where $r = |\boldsymbol{r}|$, and $\hat{r}_{\alpha}$ are the components of the vector $\boldsymbol{\hat{r}}$ = $\boldsymbol{r}$/r. These coupled equations can be rewritten in matrix-vector form:
\begin{equation}
\dot{\boldsymbol{a}}=\underline{\boldsymbol{M}}\boldsymbol{a} - i\frac{d}{\hbar}\boldsymbol{E}
\end{equation}
and the steady state solution ($\dot{\boldsymbol{a}}=0$) may be obtained by inverting a symmetric $3N\times3N$ matrix for systems where the atoms' energy levels are degenerate in $M_{J}$, and a $N\times N$ matrix for the system with Zeeman splitting. For these calculations, we average over $1.2\times 10^{5}$/N randomly distributed atom positions for the line-shape calculations and $9.6\times 10^{5}$/N randomly distributed positions for the excitation distribution calculations. The atoms are  treated as stationary and distributed according to a Gaussian density distribution:

\begin{equation}
\rho(\boldsymbol{r}) = \frac{N}{\sigma^{3}(2\pi)^{3/2}}\exp \Big(\frac{-1}{2\sigma^{2}} \big\{ \xi(y^{2} + z^{2}) + \frac{x^{2}}{\xi^{2}} \big\} \Big)
\label{eq:dens}
\end{equation}
where $\sigma$ is the standard deviation of a spherically symmetric cloud chosen to produce a specific average density $\bar{\rho}=N/(4\pi \sigma^{2})^{3/2}$ and $\xi$ is a constant that parametrizes the shape of the cloud with respect to the laser direction. Here the laser is set so that it is propagating in the $\boldsymbol{\hat{x}}$ direction and polarized in the $\boldsymbol{\hat{z}}$ direction. For spherically symmetric calculations ($\xi = 1$), we find that the results using a Gaussian density distribution agree with a constant density distribution to within $~5 \%$, for up to $10^{4}$ atoms.

 In order to avoid the usual $N^{3}$ scaling of the computation time, we solve this matrix equation using an adaptation of the Gauss-Seidel \cite{burden2011} iteration algorithm. This may be written in the form:

\begin{equation}
\boldsymbol{a}_{n+1} =\underline{\boldsymbol{D}}^{-1}\Big( \frac{id}{\hbar}\boldsymbol{E} - (\underline{\boldsymbol{M}}-\underline{\boldsymbol{D}})\boldsymbol{a}_{n}\Big)
\end{equation}
For the usual Gauss-Seidel iteration routine, $\underline{\boldsymbol{D}}$ would consist of the diagonal elements of $\underline{\boldsymbol{M}}$. However, this diverges whenever $\underline{\boldsymbol{D}}^{-1}(\underline{\boldsymbol{M}}-\underline{\boldsymbol{D}})$ contains an eigenvalue with an absolute value greater than 1. We avoid this problem by allowing $\underline{\boldsymbol{D}}$ to change depending on the current row during a matrix multiply. When a row corresponding to a particular atom is being updated in $\boldsymbol{a}$, we choose $\underline{\boldsymbol{D}}$ such that it contains current atom's couplings to its $m$ closest neighboring atoms. This routine allows the largest couplings, that would lead to the divergence discussed above to be conducted exactly, while the majority of the smaller couplings are iterated. This algorithm scales at a rate close to $N^2$, allowing us to simulate much larger numbers of atoms than we could otherwise. Note that this numerical method is much more efficient at lower densities.

\section{Single-Photon Model for a Non-Symmetric Gaussian Cloud}\label{sec:photon}
This study is conducted in the limit of weak laser intensity. Because of this, the single photon wave-function model \cite{scully2007,svidzinsky2008} should be a fair approximation. Following this model, we write our wave-function as:

\begin{eqnarray}
\ket{\Psi(t)}&=\nonumber&\beta_{+}(t)\ket{+}\ket{\{ 0\}} + \\ &\sum&_{\alpha}\beta_{\alpha}(t)\ket{-_{(\alpha)}}\ket{\{0\}} + \\   &\sum&_{\boldsymbol{k}\lambda}\gamma_{\boldsymbol{k}\lambda}(t)\ket{g}\ket{\{n_{\boldsymbol{k}\lambda}=1\}}
\end{eqnarray}
where $\ket{+}$ is the superradiant Dicke timed state:
\begin{equation}
\ket{+}=\frac{1}{\sqrt[]{N}}\sum_{j}e^{i\boldsymbol{k}_{0}\cdot\boldsymbol{r}_{j}}\ket{j}
\end{equation}
where $\ket{j}$ refers to the state where the $j^{th}$ atom is excited and the rest are in the ground state, and $\ket{-_{(\alpha)}}$ is the $\alpha^{th}$ subradiant state defined so that it is orthogonal to $\ket{+}$.

For the purpose of simplicity, we assume the cloud consists of two-level atoms polarized in the $\hat{x}$ direction driven by a laser propagating in the $\hat{z}$ direction. This differs the rest of the paper, but it simplifies the analytic calculation without changing any effects. We model the Hamiltonian as:
\begin{equation}
H = H_{0} + \hbar\sum_{j}\sum_{\boldsymbol{k}\lambda}(\hat{\epsilon}_{\boldsymbol{k}\lambda}\cdot\hat{x})g_{k}\pi^{\dagger}_{j}\hat{a}_{\boldsymbol{k}\lambda}e^{-i\omega_{k}t + i\boldsymbol{k}\cdot\boldsymbol{r}_{j}} + c.c.
\end{equation}
where $H_{0}$ is the single atom Hamiltonian, $g_{k}=-i\wp\sqrt[]{2\pi\hbar\omega_{k}/V}$ is the atom-photon coupling constant for the $\boldsymbol{k}\lambda$ mode, where $\wp$ is the dipole matrix element, $\pi^{\dagger}_{j}$ is the raising operator for atom j, and $\hat{a}_{\boldsymbol{k}\lambda}$ is the  photon lowering operator for the $\boldsymbol{k}\lambda$ mode. Switching into the Dirac picture, if we multiply the Schodinger Equation by $\bra{+}$, we obtain an equation for $\dot{\beta}_{+}(t)$:
\begin{equation}
\dot{\beta}_{+}(t) = \frac{-i}{\sqrt[]{N}}\sum_{j}\sum_{\boldsymbol{k}\lambda}(\hat{\epsilon}_{\boldsymbol{k}\lambda}\cdot\hat{x})g_{k}e^{i(\boldsymbol{k}-\boldsymbol{k}_{0})\cdot\boldsymbol{r}_{j}+i\Delta t}\gamma_{\boldsymbol{k}\lambda}(t)
\end{equation}
where $\Delta\equiv (\omega_{0}-\omega_{k})$. Similarly we obtain:

\begin{eqnarray}
\dot{\gamma}_{\boldsymbol{k}\lambda}&=\nonumber&\frac{-i}{\sqrt[]{N}}\sum_{j}\Big\{(\hat{\epsilon}_{\boldsymbol{k}\lambda}\cdot\hat{x})g^{*}_{k}e^{-i(\boldsymbol{k}-\boldsymbol{k}_{0})\cdot(\boldsymbol{r}_{j})-i\Delta t}\beta_{+}(t) \\ &+& \sum_{\alpha}c_{\alpha,j}\beta_{\alpha}(t)\Big\}
\end{eqnarray}
\begin{equation}
\dot{\beta}_{(\alpha)}(t) = \frac{-i}{\sqrt[]{N}}\sum_{j}\sum_{\boldsymbol{k}\lambda}(\hat{\epsilon}_{\boldsymbol{k}\lambda}\cdot\hat{x})g_{k}c_{\alpha,j}e^{i(\boldsymbol{k}-\boldsymbol{k}_{0})\cdot\boldsymbol{r}_{j}+i\Delta t}\gamma_{\boldsymbol{k}\lambda}(t)
\end{equation}
where $c_{\alpha,j}$ refers to amplitude of the $j^{th}$ ket of the $\alpha^{th}$ subradiant state. However, in the limit $k\sigma_{x,y,z} \gg 1$ and $N \gg 1$, there is no Agarwal-Fano coupling \cite{svidzinsky2008}, meaning that the (N-1) subradiant states will not interact with the superradiant or ground states. This allows one to write the differential equation for $\dot{\beta}_{+}(t)$ as:

\begin{eqnarray}\label{eqn:main}
\dot{\beta}_{+}(t) &=\nonumber& \frac{-1}{N}\sum_{j,j^{\prime}}\sum_{\boldsymbol{k}\lambda}\int_{0}^{t}dt^{\prime}\beta_{+}(t^{\prime})e^{i\Delta(t-t^{\prime})}(\hat{\epsilon}_{\boldsymbol{k}\lambda}\cdot\hat{x})^{2}|g_{k}|^{2} \\ &e&^{i(\boldsymbol{k}-\boldsymbol{k}_{0})\cdot(\boldsymbol{r}_{j}-\boldsymbol{r}_{j^{\prime}})}.
\end{eqnarray}
The sum over atom positions now gives:
\begin{equation}\label{eqn:sum}
\sum_{j,j^{\prime}}e^{i(\boldsymbol{k}-\boldsymbol{k}_{0})\cdot(\boldsymbol{r}_{j}-\boldsymbol{r}_{j^{\prime}})} = N + \sum_{j,j^{\prime}\ne j}e^{i(\boldsymbol{k}-\boldsymbol{k}_{0})\cdot(\boldsymbol{r}_{j}-\boldsymbol{r}_{j^{\prime}} )}
.
\end{equation}
Changing the sum  in Eq. \ref{eqn:sum} into an integral over atom positions in a Gaussian cloud elongated along the line of laser propagation gives:

\begin{eqnarray}
&N\nonumber& + N(N-1)\Big(\frac{1}{2\pi\sigma^{2}}\Big)^{3}\int\int d^{3}x_{1}d^{3}x_{2}  \\ &\exp\nonumber& \Big(\frac{-1}{2\sigma^{2}} \big\{ \xi(x_{1}^{2} + x_{2}^{2} + y_{1}^{2} + y_{2}^{2}) + \frac{z_{1}^{2} + z_{2}^{2}}{\xi^{2}} \big\} \Big) \\ &\exp& \Big(i\big\{\boldsymbol{k} - \boldsymbol{k}_{0}\big\} \cdot \big\{\boldsymbol{r}_{1} - \boldsymbol{r}_{2}\big\}  \Big).
\end{eqnarray}
Performing the integrals over atom positions gives:

\begin{eqnarray}\label{eq:pos}
&N\nonumber& + N(N-1)\exp \Big(-\sigma^{2} \\ &\big\{& \frac{1}{\xi} ( |\boldsymbol{k}_{0} - \boldsymbol{k}|_{x} + |\boldsymbol{k}_{0} - \boldsymbol{k}|_{y}) + \xi^{2}|\boldsymbol{k}_{0}-\boldsymbol{k}|_{z}\big\} \Big)
\end{eqnarray}
Setting $\boldsymbol{k}_{0} = k_{0}\hat{\boldsymbol{z}}$, and plugging Eq. (\ref{eq:pos}) into Eq. (\ref{eqn:main}) and converting the sum over $\boldsymbol{k}$ into an integral gives:

\begin{eqnarray}
\hspace{-18pt}\dot{\beta}_{+}(t) &=\frac{-V}{8\pi^{3}}\int_{0}^{\infty}\int_{0}^{\pi}\int_{0}^{2\pi} dk d\theta d\phi k^{2}\sin(\theta) \int_{0}^{t}dt^{\prime}\beta_{+}(t^{\prime})\nonumber\\ & e^{i\Delta(t-t^{\prime})}|g_{k}|^{2} (1 - \sin^{2}(\theta)\cos^{2}(\phi)) \Bigg\{ 1 + (N-1) \\ &\exp\Big(-\sigma^{2}\left\{ \frac{k^{2}\sin^{2}(\theta)}{\xi} + \xi^{2}(k_{0} - k\cos(\theta))^{2} \right\}\Big) \Bigg\}\nonumber
\end{eqnarray}
Performing the Markovian approximation, evaluating the angular integrals, and then keeping only the first order terms when invoking the limits $\xi\sim1$ and $k\sigma \gg 1$ gives:
\begin{eqnarray}
\dot{\beta}_{+}(t) &=& -\frac{\Gamma}{2}\Big(1 + \frac{\xi b_{0}}{8}\Big)\beta_{+}(t),
\end{eqnarray}
where $\Gamma = \frac{4 \omega^{3}|\wp|^{2}}{3 \hbar c^{3}}$, is the single atom decay rate and $b_{0} = \frac{3(N-1)}{k^{2}\sigma^{2}}$ is the cooperativity parameter of the gas.

\section{Numerical Results}\label{sec:results}

\subsection{Line-Broadening}

Our numerical calculations show that for spherically symmetric clouds, our the analytic prediction of Sec. III is correct. This section shows that a spherically symmetric cloud ($\xi = 1$) with a density such that $\bar{\rho}k^{3}\ll 1$, emits light with a line-width that is approximately $(1 + \zeta b_0)\Gamma$, where $\zeta$ is a numerically determined constant. Using numbers for the $^{1}S_{1}\rightarrow$ $^{3}P_{1}$ transition of $^{88}$Sr, we calculate the value of $\zeta$ for two cases: first, for the transition from the $^{3}P_{1}$ state without a magnetic field, and second for the transition from the $\boldsymbol{\hat{z}}$ polarized state, where $^{3}P_{1}$ has been Zeeman split. For the Zeeman split case, we choose the $\ket{J_{g}=0,M_{g}=0}$ to $\ket{J_{e}=1,M_{e}=0}$ transition with the driving laser polarized in the $\boldsymbol{\hat{z}}$ direction. This choice gives a maximum laser coupling between the states and gives a large effect from the dipole-dipole interaction. We show that the fractional change in line-width, $(\Gamma^{\prime}-\Gamma)/\Gamma$, is $\zeta b_{0}$ using up to $3\times 10^{3}$ atoms for the system that is degenerate in $M_{J}$ and up to $2\times 10^{4}$ atoms for the system with Zeeman splitting. For atoms experiencing no magnetic field, it is determined that $\zeta$ $\simeq$ 0.126 and for the Zeeman split case, $\zeta\simeq$  0.127. This is in good agreement with the result of $\zeta = 1/8$ derived in Sec. III using our extension of the single photon model. The fact that the numerical calculations with and without Zeeman splitting give essentially the same answer, even though one includes four states and the other only two states, is due to the strong forward dependence of the coherent interactions. Since a laser polarized in the $\boldsymbol{\hat{z}}$ direction will only illuminate the polarization state corresponding to $\boldsymbol{\hat{z}}$, and scattered radiation in the forward direction has the same polarization as the laser, coherent radiation will mainly interact with states that are $\boldsymbol{\hat{z}}$ polarized. Therefore, the states that coherently contribute to the line-broadening of the cloud will be the same for both systems. 

\begin{figure}[h]
	\includegraphics[width=0.4\textwidth]{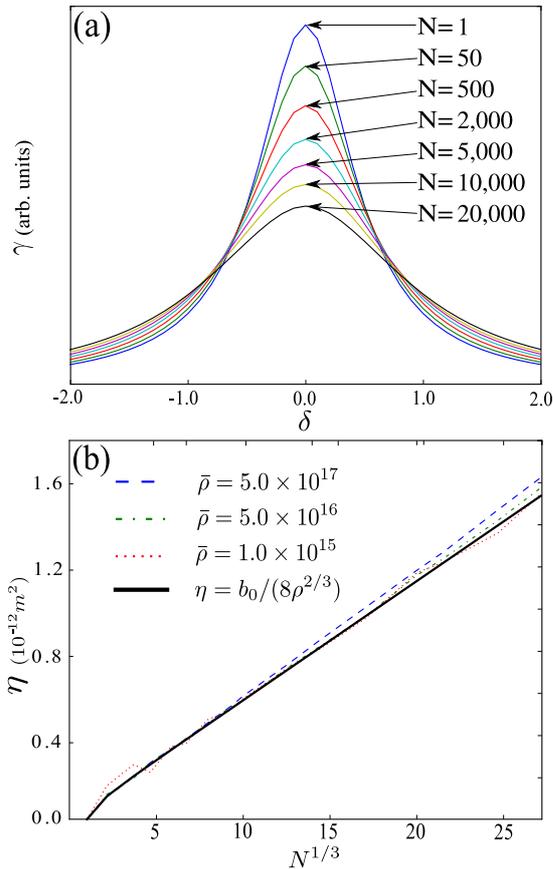}
    \caption{(a) Photon scattering rate $\gamma$ in arbitrary units versus detuning divided by $\Gamma$ ($\delta\equiv \Delta/\Gamma$), for a symmetric Gaussian cloud with $\bar{\rho}$=$5\times 10^{17}m^{-3}$. (b) Broadening normalized by $\bar{\rho}^{2/3}$, $\eta \equiv (\Gamma^{\prime}-\Gamma)/(\Gamma\rho^{2/3})$, in units of $10^{-12}m^{2}$ for densities $10^{15}m^{-3}-5.0\times 10^{17}m^{-3}$ compared to the broadening predicted using the single photon wavefunction model. These densities span a range $\bar{\rho}/k^{3}= 1.3\times 10^{-6}-6.6\times 10^{-4}$.}
    \label{figure:regular}
\end{figure}

For illustrative purposes, we rewrite our equation for a system's fractional change in line-width as, $(\Gamma^{\prime}-\Gamma)/\Gamma = 12\pi \xi\zeta \bar{\rho}^{2/3}N^{1/3}/k^2$ = $24\pi^{3/2}\xi\zeta\bar{\rho}\sigma/k^{2}$, in the limit $N \gg 1$. These equations explicitly show the parametric dependence of $\Gamma^{\prime}$. In order to demonstrate this in our calculations, Fig. \ref{figure:regular}a shows the photon scattering rate $\gamma$ versus N at a constant density of $\bar{\rho}$=$5\times 10^{17}m^{-3}$ ($\bar{\rho}/k^{3}\simeq 6.6\times 10^{-4}$). In Fig. \ref{figure:regular}b, we show that $(\Gamma^{\prime}-\Gamma)/(\Gamma\bar{\rho}^{2/3})$ when plotted against $N^{1/3}$ gives a straight line, which agrees with our analytic result for values of $\bar{\rho}$ between $10^{15}m^{-3}$ and $5\times 10^{17}m^{-3}$. This plot shows no cutoff for clouds up to $2\times 10^{4}$ atoms.

The power law $N^{1/3}$ scaling of the fractional change in line-width can be understood qualitatively using the following picture. Since the average spacing of the atoms for the largest densities shown is over $1.8\lambda$, large-scale effects only occur if there is constructive interference in the sum over \textit{many} photon-propagation paths. Here constructive interference occurs because the phase difference between the driving laser at atom $\alpha$ and at atom $\beta$ is exactly the phase difference a photon will gain when traveling from atom $\alpha$ to $\beta$ if $\frac{\boldsymbol{r}_{\alpha}-\boldsymbol{r}_{\beta}}{|\boldsymbol{r}_{\alpha}-\boldsymbol{r}_{\beta}|}\cdot\boldsymbol{\hat{k}}\simeq +1$. On the other hand if $\frac{\boldsymbol{r}_{\alpha}-\boldsymbol{r}_{\beta}}{|\boldsymbol{r}_{\alpha}-\boldsymbol{r}_{\beta}|}\cdot\boldsymbol{\hat{k}}\neq +1$, the phases will randomize, resulting in no constructive interference. Just as in the case of coherent forward scattering, where the scattered emission from a cloud adds coherently along $\boldsymbol{\hat{k}}$ \cite{rouabah2014,svidzinsky2008}, the atom-atom interactions that contribute to broadening also add coherently along $\boldsymbol{\hat{k}}$. Thus, for a given $\bar{\rho}$, the number of atoms that interact coherently and contribute to the line-broadening increases with the size of the cloud. 

\subsection{Comparison with Ref. \cite{ido2005}}

This equation for $\Gamma^{\prime}$ may be relevant to the density dependent line-broadening observed in Ref. \cite{ido2005}. For densities of $5.0\times 10^{17}m^{-3}$, Ref. \cite{ido2005} measures the homogeneous line-width of their sample to be $\simeq 29$ kHz, compared to the measured low density line-width $\simeq 14.5$ kHz, making the density dependent broadening $\simeq 14.5$ kHz. Our equation for line-broadening predicts this value to be $\simeq 26.8$ kHz, assuming a single atom lifetime of $21.3\mu $s \cite{drozdowski1997}. This calculation also assumes $N=10^6$ and a spherically symmetric cloud, both of which were not specified in Ref. \cite{ido2005}. Reference \cite{ido2005} does note that at constant $\bar{\rho}$, clouds containing more atoms have larger homogeneous line-widths, which was explained by noting that the collisional scattering lengths for atom-atom collisions increases as the relative motion between the molecules decreases. Alternatively, the increased broadening with N seen in Ref. \cite{ido2005}, could be explained using the above formula in which the fractional change in line-width increases proportional to $N^{1/3}$ for a given average density. The difference between our extrapolated result and Ref. \cite{ido2005} might be due to the simplicity of our model and/or the lack of important information about experimental parameters (for example the number of atoms in particular measurements). A more accurate calculation for this system will require the effects of atomic collisions and non-stationary atoms as well as knowledge of the shape of the atomic cloud. We address the importance of the cloud shape below.

\subsection{Dependence of Line-Shape on the Aspect Ratio}

\begin{figure}[!h]
	\includegraphics[width=0.45\textwidth,left]{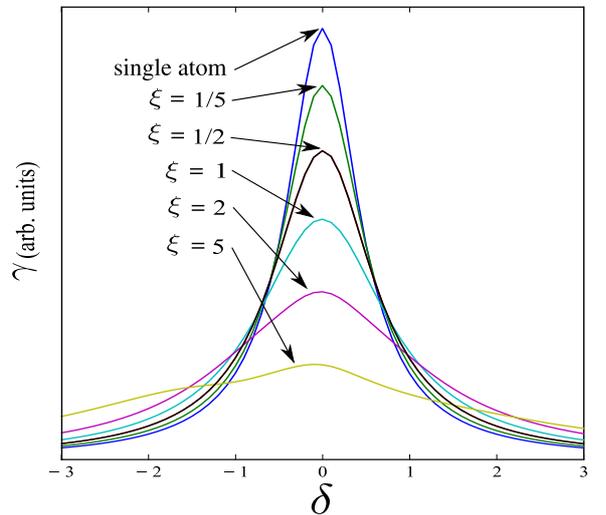}
	\caption{Photon emission rate $\gamma$ in arbitrary units versus $\delta$ (see Fig. \ref{figure:regular}) for an axially symmetric Gaussian cloud of $10^{4}$ atoms at $\bar{\rho}$ = $5\times 10^{17}m^{-3}$ for different values of $\xi$. All calculations are for the same intensity laser.}
	\label{fig:contort}
\end{figure}

Because the dipole-dipole interactions parallel to $\boldsymbol{\hat{k}}$ add constructively, the scattering rate will also depend on the shape of the atomic cloud. A cloud that is highly elongated along $\boldsymbol{\hat{k}}$ will have a larger fraction of its atoms interact constructively, causing the absorption line to be significantly more broadened. We illustrate this in Fig. \ref{fig:contort}, where we show how the Lorentzian line-shape, changes significantly as we morph the cloud from being flattened against $\boldsymbol{\hat{k}}$ to being elongated along $\boldsymbol{\hat{k}}$. In these calculations, we parametrize the spatial distribution with the variable $\xi$ defined by Eq. \ref{eq:dens}, so that the $\bar{\rho}$ is kept constant while the cloud is elongated parallel or perpendicular to $\boldsymbol{\hat{x}}$. Here the length to width ratio is proportional to $\xi^{3/2}$. 

Varying the value of $\xi$ produces several effects. First, the line-width of our scattered emission profile grows as the cloud is elongated along $\boldsymbol{\hat{k}}$ until finally it deviates from a Lorentzian profile. Numerically, we find that $\Gamma^{\prime}=(1 + \xi\zeta b_{0})\Gamma$ for  $0.0 < \xi \leq 2.0$, which agrees with our analytic result $(1 + \frac{\xi b_{0}}{8})\Gamma$. For values of $\xi > 2$, the line-shape of the scattered light begins to deviate from a Lorentzian profile. Both of these effects are shown in Fig. \ref{fig:contort}. The deviation from a Lorentzian at large values of $\xi$ implies that the approximation of 1 superradiant state and (N-1) non-interacting subradiant states \cite{scully2007,svidzinsky2008, svidzinsky_chang2008} is not valid in this regime, which means that many different modes, each with their own decay rate, begin to contribute to the overall line-shape \cite{kaiser2015}. Also, a fitted Lorentzian profile begins to show a red-shift in the peak position that increases with the value of $\xi$. These calculations show that one must consider the shape of the ensemble as a whole when calculating the line-shape of a cold, atom cloud.The linear dependence of the line-width on $\xi$ could be beneficial in experiments where the ability to control homogeneous broadening in a cloud is important.

\subsection{Counterintuitive Excitation Distribution}
\begin{figure}
	\includegraphics[width=0.43\textwidth]{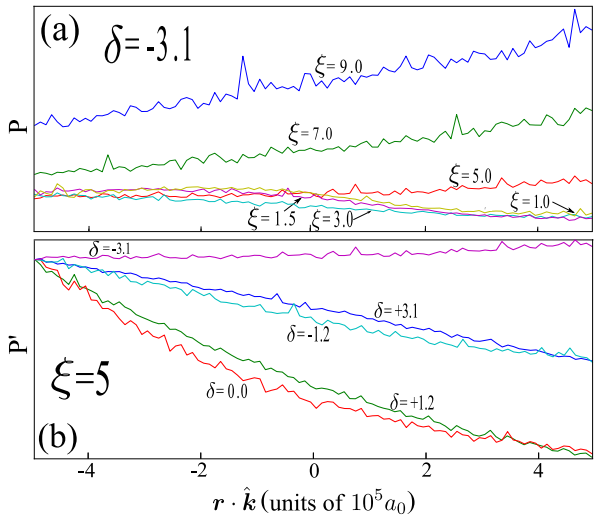}
	\caption{The average probability of an individual atom being excited in a $10^{4}$ atom cloud for various values of $\delta$ (see Fig. \ref{figure:regular}) and shapes $\xi$ versus $\boldsymbol{\hat{k}}\cdot\boldsymbol{r}$ in units of $10^{5}a_{0}$ (setting the center of the cloud as the origin). (a) The average probability of excitation (P) for a cloud driven by a red detuned laser ($\delta$=-3.1) for increasing values of $\xi$. It should be noted that in the figure, plots with larger values of $\xi$ average over fewer atoms since those clouds are stretched beyond the region plotted. (b) The `Normalized' average probability of excitation (P$^{\prime}$) for a cloud at $\xi=5$ for positive and negative laser detunings, where `Normalized' means that each curve is divided by it's initial point.}
    \label{fig:excited}
\end{figure}

Figure \ref{fig:excited} shows the average probability that an atom is excited in a cloud of $10^{4}$ atoms versus the distance along the laser direction for different values of $\xi$ (Fig. \ref{fig:excited}a) and for different detunings (Fig. \ref{fig:excited}b). For values of $\xi > 5$ and red detunings, the excitation distribution shifts towards the back of the cloud, in contrast to the exponential decay predicted by the Beer-Lambert law. A similar effect has been observed in arrays of several metallic nanospheres \cite{hernandez2005} and is due to the constructive buildup of electric field along the line of radiators. For clouds stretched parallel to $\boldsymbol{\hat{k}}$, we see that for red detuned lasers the excitation distribution of the sample is altered such that a larger fraction of the atoms are excited, causing an effective red-shift of the scattered line-shape.

\section{Discussion/Conclusion}
All of these phenomena should be observable experimentally. The calculations in Figs 2-4 focus on the $^{88}$Sr $^{1}S_{1}\rightarrow$ $^{3}P_1$ transition, but the effects shown here are independent of the value of $\Gamma$ since the damping and coupling terms in Eq. \ref{eq:a} both contain $\Gamma$. The predictions described above should be realizable for any azimuthally symmetric cloud of atoms with densities similar to those shown here when the inhomogeneous broadening is negligible. Manipulating cold atomic clouds is common in experiments \cite{ido2005, valado2013, chalony2011, marechal999, bender2010}, which makes testing the predictions of this letter feasible. For example using a magneto-optical trap (MOT), one could create a cold atomic cloud with a Gaussian density distribution equivalent to that in Eq. \ref{eq:dens} and change the values of $\sigma_{x,y,z}$ \cite{valado2013}. In this manner, the $\Gamma^{\prime} = (1 + \xi\zeta b_{0})\Gamma$ relationship could be measured by driving the cloud with a low intensity laser and measuring the emitted spectra for various numbers of atoms and cloud geometries. The predicted excitation distribution could also be shown by measuring the intensity of light emitted by different parts of the cloud.

In conclusion, collective effects can be manipulated in a cold gas of atoms, while keeping the average density of the atoms constant. When $\bar{\rho}/k^{3} \ll 1$, the fractional change in the line-width of a uniformly driven cloud of atoms undergoing $J=0$ to $J=1$ transitions is $\simeq \xi\zeta b_{0}$, where $\zeta \simeq 1/8$ is a numerically determined constant, $b_{0}$ is the cooperativity parameter $3(N-1)/(k\sigma)^{2}$, and $\xi$ is the shape parameter defined in Eq. \ref{eq:dens}. This supports the dependence on $b_{0}$ noted by previous authors using single photon wavefunction theories \cite{svidzinsky2008, svidzinsky_chang2008} as well as provide a more quantitatively accurate broadening rate. Because of the strong directionality of the coherent interactions, the photon scattering rate and the excitation distribution become strongly dependent on the shape of the cloud with respect to $\boldsymbol{\hat{k}}$. For clouds highly elongated along $\boldsymbol{\hat{k}}$, a \textit{counter-intuitive} reversal of the excitation distribution of the atoms develops, where atoms in the back of the cloud have the largest excitation probability. These calculations show that extremely small interactions can build constructively over an entire ensemble to give a strikingly large effect. The incorporation of new parameters such as collisions, inhomogeneous broadening \cite{yelin2012}, and the resultant role of subradiant states \cite{kaiser2015}, will surely provide new insights into similar systems in the future.

We thank C.H. Greene for several informative discussions. We also thank the groups of M.J. Holland, A.M. Rey, and J. Ye at Colorado for very informative poster and discussion at DAMOP 2015. This material is based upon work supported by the
National Science Foundation under Grant No. 1404419-PHY.

\bibliography{bibtex.bib}

\begin{thebibliography}{35}%
\makeatletter
\providecommand \@ifxundefined [1]{%
 \@ifx{#1\undefined}
}%
\providecommand \@ifnum [1]{%
 \ifnum #1\expandafter \@firstoftwo
 \else \expandafter \@secondoftwo
 \fi
}%
\providecommand \@ifx [1]{%
 \ifx #1\expandafter \@firstoftwo
 \else \expandafter \@secondoftwo
 \fi
}%
\providecommand \natexlab [1]{#1}%
\providecommand \enquote  [1]{``#1''}%
\providecommand \bibnamefont  [1]{#1}%
\providecommand \bibfnamefont [1]{#1}%
\providecommand \citenamefont [1]{#1}%
\providecommand \href@noop [0]{\@secondoftwo}%
\providecommand \href [0]{\begingroup \@sanitize@url \@href}%
\providecommand \@href[1]{\@@startlink{#1}\@@href}%
\providecommand \@@href[1]{\endgroup#1\@@endlink}%
\providecommand \@sanitize@url [0]{\catcode `\\12\catcode `\$12\catcode
  `\&12\catcode `\#12\catcode `\^12\catcode `\_12\catcode `\%12\relax}%
\providecommand \@@startlink[1]{}%
\providecommand \@@endlink[0]{}%
\providecommand \url  [0]{\begingroup\@sanitize@url \@url }%
\providecommand \@url [1]{\endgroup\@href {#1}{\urlprefix }}%
\providecommand \urlprefix  [0]{URL }%
\providecommand \Eprint [0]{\href }%
\providecommand \doibase [0]{http://dx.doi.org/}%
\providecommand \selectlanguage [0]{\@gobble}%
\providecommand \bibinfo  [0]{\@secondoftwo}%
\providecommand \bibfield  [0]{\@secondoftwo}%
\providecommand \translation [1]{[#1]}%
\providecommand \BibitemOpen [0]{}%
\providecommand \bibitemStop [0]{}%
\providecommand \bibitemNoStop [0]{.\EOS\space}%
\providecommand \EOS [0]{\spacefactor3000\relax}%
\providecommand \BibitemShut  [1]{\csname bibitem#1\endcsname}%
\let\auto@bib@innerbib\@empty
\bibitem [{\citenamefont {Dicke}(1954)}]{dicke1954}%
  \BibitemOpen
  \bibfield  {author} {\bibinfo {author} {\bibfnamefont {R.~H.}\ \bibnamefont
  {Dicke}},\ }\href@noop {} {\bibfield  {journal} {\bibinfo  {journal} {Phys.
  Rev.}\ }\textbf {\bibinfo {volume} {93}},\ \bibinfo {pages} {99} (\bibinfo
  {year} {1954})}\BibitemShut {NoStop}%
\bibitem [{\citenamefont {Rouabah}\ \emph {et~al.}(2014)\citenamefont
  {Rouabah}, \citenamefont {Samoylova}, \citenamefont {Bachelard},
  \citenamefont {Courteille}, \citenamefont {Kaiser},\ and\ \citenamefont
  {Piovella}}]{rouabah2014}%
  \BibitemOpen
  \bibfield  {author} {\bibinfo {author} {\bibfnamefont {M.-T.}\ \bibnamefont
  {Rouabah}}, \bibinfo {author} {\bibfnamefont {M.}~\bibnamefont {Samoylova}},
  \bibinfo {author} {\bibfnamefont {R.}~\bibnamefont {Bachelard}}, \bibinfo
  {author} {\bibfnamefont {P.~W.}\ \bibnamefont {Courteille}}, \bibinfo
  {author} {\bibfnamefont {R.}~\bibnamefont {Kaiser}}, \ and\ \bibinfo {author}
  {\bibfnamefont {N.}~\bibnamefont {Piovella}},\ }\href@noop {} {\bibfield
  {journal} {\bibinfo  {journal} {J. Opt. Soc. Am. A}\ }\textbf {\bibinfo
  {volume} {31}},\ \bibinfo {pages} {1031} (\bibinfo {year}
  {2014})}\BibitemShut {NoStop}%
\bibitem [{\citenamefont {Svidzinsky}\ \emph {et~al.}(2008)\citenamefont
  {Svidzinsky}, \citenamefont {Chang},\ and\ \citenamefont
  {Scully}}]{svidzinsky2008}%
  \BibitemOpen
  \bibfield  {author} {\bibinfo {author} {\bibfnamefont {A.~A.}\ \bibnamefont
  {Svidzinsky}}, \bibinfo {author} {\bibfnamefont {J.-T.}\ \bibnamefont
  {Chang}}, \ and\ \bibinfo {author} {\bibfnamefont {M.~O.}\ \bibnamefont
  {Scully}},\ }\href@noop {} {\bibfield  {journal} {\bibinfo  {journal} {Phys.
  Rev. Lett.}\ }\textbf {\bibinfo {volume} {100}},\ \bibinfo {pages} {160504}
  (\bibinfo {year} {2008})}\BibitemShut {NoStop}%
\bibitem [{\citenamefont {Meir}\ \emph {et~al.}(2014)\citenamefont {Meir},
  \citenamefont {Schwartz}, \citenamefont {Shahmoon}, \citenamefont {Oron},\
  and\ \citenamefont {Ozeri}}]{meir2014}%
  \BibitemOpen
  \bibfield  {author} {\bibinfo {author} {\bibfnamefont {Z.}~\bibnamefont
  {Meir}}, \bibinfo {author} {\bibfnamefont {O.}~\bibnamefont {Schwartz}},
  \bibinfo {author} {\bibfnamefont {E.}~\bibnamefont {Shahmoon}}, \bibinfo
  {author} {\bibfnamefont {D.}~\bibnamefont {Oron}}, \ and\ \bibinfo {author}
  {\bibfnamefont {R.}~\bibnamefont {Ozeri}},\ }\href@noop {} {\bibfield
  {journal} {\bibinfo  {journal} {Phys. Rev. Lett.}\ }\textbf {\bibinfo
  {volume} {113}},\ \bibinfo {pages} {193002} (\bibinfo {year}
  {2014})}\BibitemShut {NoStop}%
\bibitem [{\citenamefont {Yavuz}(2014)}]{yavuz2015}%
  \BibitemOpen
  \bibfield  {author} {\bibinfo {author} {\bibfnamefont {D.~D.}\ \bibnamefont
  {Yavuz}},\ }\href {\doibase 10.1364/JOSAB.31.002665} {\bibfield  {journal}
  {\bibinfo  {journal} {J. Opt. Soc. Am. B}\ }\textbf {\bibinfo {volume}
  {31}},\ \bibinfo {pages} {2665} (\bibinfo {year} {2014})}\BibitemShut
  {NoStop}%
\bibitem [{\citenamefont {Smith}\ and\ \citenamefont
  {Burnett}(1991)}]{burnett1991}%
  \BibitemOpen
  \bibfield  {author} {\bibinfo {author} {\bibfnamefont {A.~M.}\ \bibnamefont
  {Smith}}\ and\ \bibinfo {author} {\bibfnamefont {K.}~\bibnamefont
  {Burnett}},\ }\href {\doibase 10.1364/JOSAB.8.001592} {\bibfield  {journal}
  {\bibinfo  {journal} {J. Opt. Soc. Am. B}\ }\textbf {\bibinfo {volume} {8}},\
  \bibinfo {pages} {1592} (\bibinfo {year} {1991})}\BibitemShut {NoStop}%
\bibitem [{\citenamefont {Pellegrino}\ \emph {et~al.}(2014)\citenamefont
  {Pellegrino}, \citenamefont {Bourgain}, \citenamefont {Jennewein},
  \citenamefont {Sortais}, \citenamefont {Browaeys}, \citenamefont {Jenkins},\
  and\ \citenamefont {Ruostekoski}}]{pellegrino2014}%
  \BibitemOpen
  \bibfield  {author} {\bibinfo {author} {\bibfnamefont {J.}~\bibnamefont
  {Pellegrino}}, \bibinfo {author} {\bibfnamefont {R.}~\bibnamefont
  {Bourgain}}, \bibinfo {author} {\bibfnamefont {S.}~\bibnamefont {Jennewein}},
  \bibinfo {author} {\bibfnamefont {Y.~R.}\ \bibnamefont {Sortais}}, \bibinfo
  {author} {\bibfnamefont {A.}~\bibnamefont {Browaeys}}, \bibinfo {author}
  {\bibfnamefont {S.}~\bibnamefont {Jenkins}}, \ and\ \bibinfo {author}
  {\bibfnamefont {J.}~\bibnamefont {Ruostekoski}},\ }\href@noop {} {\bibfield
  {journal} {\bibinfo  {journal} {Phys. Rev. Lett.}\ }\textbf {\bibinfo
  {volume} {113}},\ \bibinfo {pages} {133602} (\bibinfo {year}
  {2014})}\BibitemShut {NoStop}%
\bibitem [{\citenamefont {Ido}\ \emph {et~al.}(2005)\citenamefont {Ido},
  \citenamefont {Loftus}, \citenamefont {Boyd}, \citenamefont {Ludlow},
  \citenamefont {Holman},\ and\ \citenamefont {Ye}}]{ido2005}%
  \BibitemOpen
  \bibfield  {author} {\bibinfo {author} {\bibfnamefont {T.}~\bibnamefont
  {Ido}}, \bibinfo {author} {\bibfnamefont {T.~H.}\ \bibnamefont {Loftus}},
  \bibinfo {author} {\bibfnamefont {M.~M.}\ \bibnamefont {Boyd}}, \bibinfo
  {author} {\bibfnamefont {A.~D.}\ \bibnamefont {Ludlow}}, \bibinfo {author}
  {\bibfnamefont {K.~W.}\ \bibnamefont {Holman}}, \ and\ \bibinfo {author}
  {\bibfnamefont {J.}~\bibnamefont {Ye}},\ }\href@noop {} {\bibfield  {journal}
  {\bibinfo  {journal} {Phys. Rev. Lett.}\ }\textbf {\bibinfo {volume} {94}},\
  \bibinfo {pages} {153001} (\bibinfo {year} {2005})}\BibitemShut {NoStop}%
\bibitem [{\citenamefont {Chang}\ \emph {et~al.}(2004)\citenamefont {Chang},
  \citenamefont {Ye},\ and\ \citenamefont {Lukin}}]{chang2004}%
  \BibitemOpen
  \bibfield  {author} {\bibinfo {author} {\bibfnamefont {D.}~\bibnamefont
  {Chang}}, \bibinfo {author} {\bibfnamefont {J.}~\bibnamefont {Ye}}, \ and\
  \bibinfo {author} {\bibfnamefont {M.}~\bibnamefont {Lukin}},\ }\href@noop {}
  {\bibfield  {journal} {\bibinfo  {journal} {Phys. Rev. A}\ }\textbf {\bibinfo
  {volume} {69}},\ \bibinfo {pages} {023810} (\bibinfo {year}
  {2004})}\BibitemShut {NoStop}%
\bibitem [{\citenamefont {Katori}\ \emph {et~al.}(2003)\citenamefont {Katori},
  \citenamefont {Takamoto}, \citenamefont {Pal'chikov},\ and\ \citenamefont
  {Ovsiannikov}}]{katori2003}%
  \BibitemOpen
  \bibfield  {author} {\bibinfo {author} {\bibfnamefont {H.}~\bibnamefont
  {Katori}}, \bibinfo {author} {\bibfnamefont {M.}~\bibnamefont {Takamoto}},
  \bibinfo {author} {\bibfnamefont {V.~G.}\ \bibnamefont {Pal'chikov}}, \ and\
  \bibinfo {author} {\bibfnamefont {V.~D.}\ \bibnamefont {Ovsiannikov}},\
  }\href {\doibase 10.1103/PhysRevLett.91.173005} {\bibfield  {journal}
  {\bibinfo  {journal} {Phys. Rev. Lett.}\ }\textbf {\bibinfo {volume} {91}},\
  \bibinfo {pages} {173005} (\bibinfo {year} {2003})}\BibitemShut {NoStop}%
\bibitem [{\citenamefont {Ludlow}\ \emph {et~al.}(2015)\citenamefont {Ludlow},
  \citenamefont {Boyd}, \citenamefont {Ye}, \citenamefont {Peik},\ and\
  \citenamefont {Schmidt}}]{ludlow2015}%
  \BibitemOpen
  \bibfield  {author} {\bibinfo {author} {\bibfnamefont {A.~D.}\ \bibnamefont
  {Ludlow}}, \bibinfo {author} {\bibfnamefont {M.~M.}\ \bibnamefont {Boyd}},
  \bibinfo {author} {\bibfnamefont {J.}~\bibnamefont {Ye}}, \bibinfo {author}
  {\bibfnamefont {E.}~\bibnamefont {Peik}}, \ and\ \bibinfo {author}
  {\bibfnamefont {P.~O.}\ \bibnamefont {Schmidt}},\ }\href {\doibase
  10.1103/RevModPhys.87.637} {\bibfield  {journal} {\bibinfo  {journal} {Rev.
  Mod. Phys.}\ }\textbf {\bibinfo {volume} {87}},\ \bibinfo {pages} {637}
  (\bibinfo {year} {2015})}\BibitemShut {NoStop}%
\bibitem [{\citenamefont {Shiga}\ \emph {et~al.}(2009)\citenamefont {Shiga},
  \citenamefont {Li}, \citenamefont {Ito}, \citenamefont {Nagano},
  \citenamefont {Ido}, \citenamefont {Bielska}, \citenamefont
  {Trawi\ifmmode~\acute{n}\else \'{n}\fi{}ski},\ and\ \citenamefont
  {Ciury\l{}o}}]{shiga2009}%
  \BibitemOpen
  \bibfield  {author} {\bibinfo {author} {\bibfnamefont {N.}~\bibnamefont
  {Shiga}}, \bibinfo {author} {\bibfnamefont {Y.}~\bibnamefont {Li}}, \bibinfo
  {author} {\bibfnamefont {H.}~\bibnamefont {Ito}}, \bibinfo {author}
  {\bibfnamefont {S.}~\bibnamefont {Nagano}}, \bibinfo {author} {\bibfnamefont
  {T.}~\bibnamefont {Ido}}, \bibinfo {author} {\bibfnamefont {K.}~\bibnamefont
  {Bielska}}, \bibinfo {author} {\bibfnamefont {R.~S.}\ \bibnamefont
  {Trawi\ifmmode~\acute{n}\else \'{n}\fi{}ski}}, \ and\ \bibinfo {author}
  {\bibfnamefont {R.}~\bibnamefont {Ciury\l{}o}},\ }\href {\doibase
  10.1103/PhysRevA.80.030501} {\bibfield  {journal} {\bibinfo  {journal} {Phys.
  Rev. A}\ }\textbf {\bibinfo {volume} {80}},\ \bibinfo {pages} {030501}
  (\bibinfo {year} {2009})}\BibitemShut {NoStop}%
\bibitem [{\citenamefont {Friedberg}\ \emph {et~al.}(1973)\citenamefont
  {Friedberg}, \citenamefont {Hartmann},\ and\ \citenamefont
  {Manassah}}]{friedberg1973}%
  \BibitemOpen
  \bibfield  {author} {\bibinfo {author} {\bibfnamefont {R.}~\bibnamefont
  {Friedberg}}, \bibinfo {author} {\bibfnamefont {S.~R.}\ \bibnamefont
  {Hartmann}}, \ and\ \bibinfo {author} {\bibfnamefont {J.~T.}\ \bibnamefont
  {Manassah}},\ }\href@noop {} {\bibfield  {journal} {\bibinfo  {journal}
  {Phys. Rep.}\ }\textbf {\bibinfo {volume} {7}},\ \bibinfo {pages} {101}
  (\bibinfo {year} {1973})}\BibitemShut {NoStop}%
\bibitem [{\citenamefont {Baranger}(1958)}]{baranger1958}%
  \BibitemOpen
  \bibfield  {author} {\bibinfo {author} {\bibfnamefont {M.}~\bibnamefont
  {Baranger}},\ }\href@noop {} {\bibfield  {journal} {\bibinfo  {journal}
  {Phys. Rev.}\ }\textbf {\bibinfo {volume} {111}},\ \bibinfo {pages} {481}
  (\bibinfo {year} {1958})}\BibitemShut {NoStop}%
\bibitem [{\citenamefont {Breene}(1970)}]{breene1970}%
  \BibitemOpen
  \bibfield  {author} {\bibinfo {author} {\bibfnamefont {R.}~\bibnamefont
  {Breene}},\ }\href@noop {} {\bibfield  {journal} {\bibinfo  {journal} {Phys.
  Lett. A}\ }\textbf {\bibinfo {volume} {32}},\ \bibinfo {pages} {466}
  (\bibinfo {year} {1970})}\BibitemShut {NoStop}%
\bibitem [{\citenamefont {Svidzinsky}\ and\ \citenamefont
  {Chang}(2008)}]{svidzinsky_chang2008}%
  \BibitemOpen
  \bibfield  {author} {\bibinfo {author} {\bibfnamefont {A.}~\bibnamefont
  {Svidzinsky}}\ and\ \bibinfo {author} {\bibfnamefont {J.-T.}\ \bibnamefont
  {Chang}},\ }\href {\doibase 10.1103/PhysRevA.77.043833} {\bibfield  {journal}
  {\bibinfo  {journal} {Phys. Rev. A}\ }\textbf {\bibinfo {volume} {77}},\
  \bibinfo {pages} {043833} (\bibinfo {year} {2008})}\BibitemShut {NoStop}%
\bibitem [{\citenamefont {Bienaim\'e}\ \emph {et~al.}(2012)\citenamefont
  {Bienaim\'e}, \citenamefont {Piovella},\ and\ \citenamefont
  {Kaiser}}]{bienaime2012}%
  \BibitemOpen
  \bibfield  {author} {\bibinfo {author} {\bibfnamefont {T.}~\bibnamefont
  {Bienaim\'e}}, \bibinfo {author} {\bibfnamefont {N.}~\bibnamefont
  {Piovella}}, \ and\ \bibinfo {author} {\bibfnamefont {R.}~\bibnamefont
  {Kaiser}},\ }\href {\doibase 10.1103/PhysRevLett.108.123602} {\bibfield
  {journal} {\bibinfo  {journal} {Phys. Rev. Lett.}\ }\textbf {\bibinfo
  {volume} {108}},\ \bibinfo {pages} {123602} (\bibinfo {year}
  {2012})}\BibitemShut {NoStop}%
\bibitem [{\citenamefont {Courteille}\ \emph {et~al.}(2010)\citenamefont
  {Courteille}, \citenamefont {Bux}, \citenamefont {Lucioni}, \citenamefont
  {Lauber}, \citenamefont {Bienaime}, \citenamefont {Kaiser},\ and\
  \citenamefont {Piovella}}]{courteille2010}%
  \BibitemOpen
  \bibfield  {author} {\bibinfo {author} {\bibfnamefont {P.~W.}\ \bibnamefont
  {Courteille}}, \bibinfo {author} {\bibfnamefont {S.}~\bibnamefont {Bux}},
  \bibinfo {author} {\bibfnamefont {E.}~\bibnamefont {Lucioni}}, \bibinfo
  {author} {\bibfnamefont {K.}~\bibnamefont {Lauber}}, \bibinfo {author}
  {\bibfnamefont {T.}~\bibnamefont {Bienaime}}, \bibinfo {author}
  {\bibfnamefont {R.}~\bibnamefont {Kaiser}}, \ and\ \bibinfo {author}
  {\bibfnamefont {N.}~\bibnamefont {Piovella}},\ }\href@noop {} {\bibfield
  {journal} {\bibinfo  {journal} {The Eur. Phys. J. D}\ }\textbf {\bibinfo
  {volume} {58}},\ \bibinfo {pages} {69} (\bibinfo {year} {2010})}\BibitemShut
  {NoStop}%
\bibitem [{\citenamefont {Scully}(2007)}]{scully2007}%
  \BibitemOpen
  \bibfield  {author} {\bibinfo {author} {\bibfnamefont {M.}~\bibnamefont
  {Scully}},\ }\href@noop {} {\bibfield  {journal} {\bibinfo  {journal} {Laser
  Phys.}\ }\textbf {\bibinfo {volume} {17}},\ \bibinfo {pages} {635} (\bibinfo
  {year} {2007})}\BibitemShut {NoStop}%
\bibitem [{\citenamefont {Javanainen}\ \emph {et~al.}(2014)\citenamefont
  {Javanainen}, \citenamefont {Ruostekoski}, \citenamefont {Li},\ and\
  \citenamefont {Yoo}}]{javanainen2014}%
  \BibitemOpen
  \bibfield  {author} {\bibinfo {author} {\bibfnamefont {J.}~\bibnamefont
  {Javanainen}}, \bibinfo {author} {\bibfnamefont {J.}~\bibnamefont
  {Ruostekoski}}, \bibinfo {author} {\bibfnamefont {Y.}~\bibnamefont {Li}}, \
  and\ \bibinfo {author} {\bibfnamefont {S.-M.}\ \bibnamefont {Yoo}},\
  }\href@noop {} {\bibfield  {journal} {\bibinfo  {journal} {Phys. Rev. Lett.}\
  }\textbf {\bibinfo {volume} {112}},\ \bibinfo {pages} {113603} (\bibinfo
  {year} {2014})}\BibitemShut {NoStop}%
\bibitem [{\citenamefont {Terhal}\ and\ \citenamefont
  {Burkard}(2005)}]{terhal2005}%
  \BibitemOpen
  \bibfield  {author} {\bibinfo {author} {\bibfnamefont {B.~M.}\ \bibnamefont
  {Terhal}}\ and\ \bibinfo {author} {\bibfnamefont {G.}~\bibnamefont
  {Burkard}},\ }\href {\doibase 10.1103/PhysRevA.71.012336} {\bibfield
  {journal} {\bibinfo  {journal} {Phys. Rev. A}\ }\textbf {\bibinfo {volume}
  {71}},\ \bibinfo {pages} {012336} (\bibinfo {year} {2005})}\BibitemShut
  {NoStop}%
\bibitem [{\citenamefont {Ng}\ and\ \citenamefont {Preskill}(2009)}]{hui2009}%
  \BibitemOpen
  \bibfield  {author} {\bibinfo {author} {\bibfnamefont {H.~K.}\ \bibnamefont
  {Ng}}\ and\ \bibinfo {author} {\bibfnamefont {J.}~\bibnamefont {Preskill}},\
  }\href {\doibase 10.1103/PhysRevA.79.032318} {\bibfield  {journal} {\bibinfo
  {journal} {Phys. Rev. A}\ }\textbf {\bibinfo {volume} {79}},\ \bibinfo
  {pages} {032318} (\bibinfo {year} {2009})}\BibitemShut {NoStop}%
\bibitem [{\citenamefont {Jenkins}\ and\ \citenamefont
  {Ruostekoski}(2012)}]{jenkins2012}%
  \BibitemOpen
  \bibfield  {author} {\bibinfo {author} {\bibfnamefont {S.~D.}\ \bibnamefont
  {Jenkins}}\ and\ \bibinfo {author} {\bibfnamefont {J.}~\bibnamefont
  {Ruostekoski}},\ }\href@noop {} {\bibfield  {journal} {\bibinfo  {journal}
  {Phys. Rev. A}\ }\textbf {\bibinfo {volume} {86}},\ \bibinfo {pages} {031602}
  (\bibinfo {year} {2012})}\BibitemShut {NoStop}%
\bibitem [{\citenamefont {Svidzinsky}\ \emph {et~al.}(2010)\citenamefont
  {Svidzinsky}, \citenamefont {Chang},\ and\ \citenamefont
  {Scully}}]{svidzinsky2010}%
  \BibitemOpen
  \bibfield  {author} {\bibinfo {author} {\bibfnamefont {A.~A.}\ \bibnamefont
  {Svidzinsky}}, \bibinfo {author} {\bibfnamefont {J.-T.}\ \bibnamefont
  {Chang}}, \ and\ \bibinfo {author} {\bibfnamefont {M.~O.}\ \bibnamefont
  {Scully}},\ }\href {\doibase 10.1103/PhysRevA.81.053821} {\bibfield
  {journal} {\bibinfo  {journal} {Phys. Rev. A}\ }\textbf {\bibinfo {volume}
  {81}},\ \bibinfo {pages} {053821} (\bibinfo {year} {2010})}\BibitemShut
  {NoStop}%
\bibitem [{\citenamefont {Ruostekoski}\ and\ \citenamefont
  {Javanainen}(1997)}]{ruostekoski1997}%
  \BibitemOpen
  \bibfield  {author} {\bibinfo {author} {\bibfnamefont {J.}~\bibnamefont
  {Ruostekoski}}\ and\ \bibinfo {author} {\bibfnamefont {J.}~\bibnamefont
  {Javanainen}},\ }\href@noop {} {\bibfield  {journal} {\bibinfo  {journal}
  {Phys. Rev. A}\ }\textbf {\bibinfo {volume} {55}},\ \bibinfo {pages} {513}
  (\bibinfo {year} {1997})}\BibitemShut {NoStop}%
\bibitem [{\citenamefont {Jackson}\ and\ \citenamefont
  {Jackson}(1999)}]{jackson1999}%
  \BibitemOpen
  \bibfield  {author} {\bibinfo {author} {\bibfnamefont {J.~D.}\ \bibnamefont
  {Jackson}}\ and\ \bibinfo {author} {\bibfnamefont {J.~D.}\ \bibnamefont
  {Jackson}},\ }\href@noop {} {\emph {\bibinfo {title} {Classical
  electrodynamics}}},\ Vol.~\bibinfo {volume} {3}\ (\bibinfo  {publisher}
  {Wiley New York etc.},\ \bibinfo {year} {1999})\BibitemShut {NoStop}%
\bibitem [{\citenamefont {Burden}\ and\ \citenamefont
  {Faires}(2011)}]{burden2011}%
  \BibitemOpen
  \bibfield  {author} {\bibinfo {author} {\bibfnamefont {R.~L.}\ \bibnamefont
  {Burden}}\ and\ \bibinfo {author} {\bibfnamefont {J.~D.}\ \bibnamefont
  {Faires}},\ }\href@noop {} {\emph {\bibinfo {title} {Numerical analysis}}},\
  Vol.~\bibinfo {volume} {9}\ (\bibinfo {year} {2011})\BibitemShut {NoStop}%
\bibitem [{\citenamefont {Drozdowski}\ \emph {et~al.}(1997)\citenamefont
  {Drozdowski}, \citenamefont {Ignaciuk}, \citenamefont {Kwela},\ and\
  \citenamefont {Heldt}}]{drozdowski1997}%
  \BibitemOpen
  \bibfield  {author} {\bibinfo {author} {\bibfnamefont {R.}~\bibnamefont
  {Drozdowski}}, \bibinfo {author} {\bibfnamefont {M.}~\bibnamefont
  {Ignaciuk}}, \bibinfo {author} {\bibfnamefont {J.}~\bibnamefont {Kwela}}, \
  and\ \bibinfo {author} {\bibfnamefont {J.}~\bibnamefont {Heldt}},\
  }\href@noop {} {\bibfield  {journal} {\bibinfo  {journal} {Zeitschrift
  f{\"u}r Physik D Atoms, Molecules and Clusters}\ }\textbf {\bibinfo {volume}
  {41}},\ \bibinfo {pages} {125} (\bibinfo {year} {1997})}\BibitemShut
  {NoStop}%
\bibitem [{\citenamefont {William~Guerin}\ and\ \citenamefont
  {Kaiser}(2015)}]{kaiser2015}%
  \BibitemOpen
  \bibfield  {author} {\bibinfo {author} {\bibfnamefont {M.~A.}\ \bibnamefont
  {William~Guerin}}\ and\ \bibinfo {author} {\bibfnamefont {R.}~\bibnamefont
  {Kaiser}},\ }\href@noop {} {\bibfield  {journal} {\bibinfo  {journal}
  {http://arxiv.org/1509.00227}\ } (\bibinfo {year} {2015})}\BibitemShut
  {NoStop}%
\bibitem [{\citenamefont {Hern{\'a}ndez}\ \emph {et~al.}(2005)\citenamefont
  {Hern{\'a}ndez}, \citenamefont {Noordam},\ and\ \citenamefont
  {Robicheaux}}]{hernandez2005}%
  \BibitemOpen
  \bibfield  {author} {\bibinfo {author} {\bibfnamefont {J.}~\bibnamefont
  {Hern{\'a}ndez}}, \bibinfo {author} {\bibfnamefont {L.}~\bibnamefont
  {Noordam}}, \ and\ \bibinfo {author} {\bibfnamefont {F.}~\bibnamefont
  {Robicheaux}},\ }\href@noop {} {\bibfield  {journal} {\bibinfo  {journal} {J.
  Phys. Chem. B}\ }\textbf {\bibinfo {volume} {109}},\ \bibinfo {pages} {15808}
  (\bibinfo {year} {2005})}\BibitemShut {NoStop}%
\bibitem [{\citenamefont {Valado}\ \emph {et~al.}(2013)\citenamefont {Valado},
  \citenamefont {Malossi}, \citenamefont {Scotto}, \citenamefont {Ciampini},
  \citenamefont {Arimondo},\ and\ \citenamefont {Morsch}}]{valado2013}%
  \BibitemOpen
  \bibfield  {author} {\bibinfo {author} {\bibfnamefont {M.~M.}\ \bibnamefont
  {Valado}}, \bibinfo {author} {\bibfnamefont {N.}~\bibnamefont {Malossi}},
  \bibinfo {author} {\bibfnamefont {S.}~\bibnamefont {Scotto}}, \bibinfo
  {author} {\bibfnamefont {D.}~\bibnamefont {Ciampini}}, \bibinfo {author}
  {\bibfnamefont {E.}~\bibnamefont {Arimondo}}, \ and\ \bibinfo {author}
  {\bibfnamefont {O.}~\bibnamefont {Morsch}},\ }\href {\doibase
  10.1103/PhysRevA.88.045401} {\bibfield  {journal} {\bibinfo  {journal} {Phys.
  Rev. A}\ }\textbf {\bibinfo {volume} {88}},\ \bibinfo {pages} {045401}
  (\bibinfo {year} {2013})}\BibitemShut {NoStop}%
\bibitem [{\citenamefont {Chalony}\ \emph {et~al.}(2011)\citenamefont
  {Chalony}, \citenamefont {Pierrat}, \citenamefont {Delande},\ and\
  \citenamefont {Wilkowski}}]{chalony2011}%
  \BibitemOpen
  \bibfield  {author} {\bibinfo {author} {\bibfnamefont {M.}~\bibnamefont
  {Chalony}}, \bibinfo {author} {\bibfnamefont {R.}~\bibnamefont {Pierrat}},
  \bibinfo {author} {\bibfnamefont {D.}~\bibnamefont {Delande}}, \ and\
  \bibinfo {author} {\bibfnamefont {D.}~\bibnamefont {Wilkowski}},\ }\href
  {\doibase 10.1103/PhysRevA.84.011401} {\bibfield  {journal} {\bibinfo
  {journal} {Phys. Rev. A}\ }\textbf {\bibinfo {volume} {84}},\ \bibinfo
  {pages} {011401} (\bibinfo {year} {2011})}\BibitemShut {NoStop}%
\bibitem [{\citenamefont {Mar\'echal}\ \emph {et~al.}(1999)\citenamefont
  {Mar\'echal}, \citenamefont {Guibal}, \citenamefont {Bossennec},
  \citenamefont {Barb\'e}, \citenamefont {Keller},\ and\ \citenamefont
  {Gorceix}}]{marechal999}%
  \BibitemOpen
  \bibfield  {author} {\bibinfo {author} {\bibfnamefont {E.}~\bibnamefont
  {Mar\'echal}}, \bibinfo {author} {\bibfnamefont {S.}~\bibnamefont {Guibal}},
  \bibinfo {author} {\bibfnamefont {J.-L.}\ \bibnamefont {Bossennec}}, \bibinfo
  {author} {\bibfnamefont {R.}~\bibnamefont {Barb\'e}}, \bibinfo {author}
  {\bibfnamefont {J.-C.}\ \bibnamefont {Keller}}, \ and\ \bibinfo {author}
  {\bibfnamefont {O.}~\bibnamefont {Gorceix}},\ }\href {\doibase
  10.1103/PhysRevA.59.4636} {\bibfield  {journal} {\bibinfo  {journal} {Phys.
  Rev. A}\ }\textbf {\bibinfo {volume} {59}},\ \bibinfo {pages} {4636}
  (\bibinfo {year} {1999})}\BibitemShut {NoStop}%
\bibitem [{\citenamefont {Bender}\ \emph {et~al.}(2010)\citenamefont {Bender},
  \citenamefont {Stehle}, \citenamefont {Slama}, \citenamefont {Kaiser},
  \citenamefont {Piovella}, \citenamefont {Zimmermann},\ and\ \citenamefont
  {Courteille}}]{bender2010}%
  \BibitemOpen
  \bibfield  {author} {\bibinfo {author} {\bibfnamefont {H.}~\bibnamefont
  {Bender}}, \bibinfo {author} {\bibfnamefont {C.}~\bibnamefont {Stehle}},
  \bibinfo {author} {\bibfnamefont {S.}~\bibnamefont {Slama}}, \bibinfo
  {author} {\bibfnamefont {R.}~\bibnamefont {Kaiser}}, \bibinfo {author}
  {\bibfnamefont {N.}~\bibnamefont {Piovella}}, \bibinfo {author}
  {\bibfnamefont {C.}~\bibnamefont {Zimmermann}}, \ and\ \bibinfo {author}
  {\bibfnamefont {P.~W.}\ \bibnamefont {Courteille}},\ }\href {\doibase
  10.1103/PhysRevA.82.011404} {\bibfield  {journal} {\bibinfo  {journal} {Phys.
  Rev. A}\ }\textbf {\bibinfo {volume} {82}},\ \bibinfo {pages} {011404}
  (\bibinfo {year} {2010})}\BibitemShut {NoStop}%
\bibitem [{\citenamefont {Lin}\ and\ \citenamefont {Yelin}(2012)}]{yelin2012}%
  \BibitemOpen
  \bibfield  {author} {\bibinfo {author} {\bibfnamefont {G.-D.}\ \bibnamefont
  {Lin}}\ and\ \bibinfo {author} {\bibfnamefont {S.~F.}\ \bibnamefont
  {Yelin}},\ }\href {\doibase 10.1103/PhysRevA.85.033831} {\bibfield  {journal}
  {\bibinfo  {journal} {Phys. Rev. A}\ }\textbf {\bibinfo {volume} {85}},\
  \bibinfo {pages} {033831} (\bibinfo {year} {2012})}\BibitemShut {NoStop}%
\end{thebibliography}%
\end{document}